\documentclass[multphys,vecphys]{svmult}

\usepackage{makeidx}         
\usepackage{graphicx}        
\usepackage{multicol}        
\usepackage[bottom]{footmisc}

\makeindex             

\begin{document}

\title*{Wakes in Dark Matter Halos}
\author{Burkhard Fuchs}
\institute{Astronomisches Rechen-Institut, M\"onchhofstr.~12-14, 69120
Heidelberg, Germany
\texttt{fuchs@ari.uni-heidelberg.de}}
%
%
\maketitle

\begin{abstract}
I discuss the dynamical interaction of galactic disks with the surrounding dark 
matter halos. In particular it is demonstrated that if the self--gravitating 
shearing sheet, a model of a patch of a galactic disk, is embedded in a live
dark halo, this has a strong effect on the dynamics of density waves in the
sheet. I describe how the density waves and the halo interact via halo particles
either on orbits in resonance with the wave or on non-resonant orbits. Contrary
to expectation the presence of the halo leads to a very considerable
enhancement of the amplitudes of the density waves in the shearing sheet. This
effect appears to be the equivalent of the recently reported enhanced growth of 
bars in numerically simulated stellar disks embedded in live dark halos.
Finally I discuss the counterparts of the perturbations of the disk in the dark
halo.
\end{abstract}

\section{Introduction}
\label{sec:1}

Dark halos are usually thought to stabilize galactic disks against 
non-axisymmetric instabilities. This was
first proposed by Ostriker \& Peebles (1973) on the basis of --
low--resolution -- numerical simulations. Their physical argument was that
the presence of a dark halo reduces the destabilizing self--gravity of the
disks. Doubts about an entirely passive role of dark halos were raised by 
Toomre (1977), but he (Toomre 1981) also pointed out that a dense core of a 
dark halo may cut the feed--back loop of the corotation amplifier of bars or
spiral density waves and suppress thus their growth. Recent high-resolution 
numerical simulations by Athanassoula (2002, 2003), also inherent in the
work of Debattista \& Sellwood (2000), have shown that quite the reverse, a 
{\em destabilization} of
disks immersed in dark halos, might be actually true. Athanassoula (2002) 
demonstrated clearly that much stronger bars grow in the simulations 
if the disk is embedded in a live dark halo instead of a static halo potential.
This is attributed to angular momentum transfer from the bar to the halo via 
halo particles on resonant orbits. Angular momentum exchange between disk and 
halo has been addressed since the pioneering work of Weinberg (1985) in many 
studies theoretically or by numerical simulations and I refer to Athanassoula 
(2003) for an overview of the literature. Toomre (1981) has shown how the bar
instability can be understood as an interference of spiral density waves in a
resonance cavity between the corotation amplifier and an inner reflector
(cf.~also Fuchs 2004b). Thus it is to be expected that a live dark halo will be 
also responsive to spiral density waves and develop wakes. That this is indeed
the case has been demonstrated by Fuchs (2004a) employing the 
shearing sheet model. This adds to the  confidence in the results of the
numerical work on bar growth in galactic disks.

The shearing sheet (Goldreich \& Lynden--Bell 1965, Julian \& Toomre 
1966) model has been developed as a tool to study the dynamics of
galactic disks and is particularly well suited to describe theoretically
the dynamical mechanisms responsible for the formation of spiral arms.
For the sake of simplicity the model describes only the dynamics of a 
patch of a galactic disk. It is assumed to be infinitesimally thin and 
its radial size is assumed to be much smaller than the disk. Polar 
coordinates can be therefore rectified to pseudo-Cartesian
coordinates and the velocity field of the differential rotation of the
disk can be approximated by a linear shear flow. These simplifications
allow an analytical treatment of the problem, which helps also in the present 
case to clarify the underlying physical processes operating in the disk.

\index{shearing sheet}

\section{Shearing sheet model}
\label{sec:2}

The basic disk model is the stellardynamical shearing sheet, which describes 
the local dynamics of a patch of a thin, differentially rotating stellar disk.
Stellar orbits are calculated in a frame at a distance $r_0$
from the galactic center, rotating with an angular velocity $\Omega_0$.
Pseudo--Cartesian coordinates $x$ and $y$ point in the radial direction and 
tangential to the direction of galactic rotation,
respectively. The differential rotation of the disk is approximated as a
parallel shear flow, $v = -2Ax$, with $A$ denoting Oort's constant.  The 
surface density $\Sigma_0$ is assumed to be constant over the entire region.
As is well known (cf.~Julian \& Toomre 1966) the
stellar orbits in this model are simply epicyclic orbits and the phase space
distribution function of the stars $f_0$,
as derived from the time-independent Boltzmann-equation,
is a Schwarzschild-distribution. A cartoon of the shearing
sheet model is shown in Fig.~1. 
\begin{figure}
\centering
%
\caption{The shearing sheet model}
\label{fig:1}       
\end{figure}

The disk is subjected to potential perturbations
\begin{equation}
\delta\Phi = \int d\omega \int dk_{\rm x} \int dk_{\rm y}
\Phi_{\rm{\bf k},\omega}
e^{i(\omega t + k_{\rm x} x + k_{\rm y} y)},
\end{equation}
and the response of the disk, $f_1$, is calculated from the linearized
Boltzmann-equation
\begin{equation}
\frac{\partial f_1}{\partial t} + [f_0,\delta\Phi] +[f_1,H_0] = 0\,,
\end{equation}
written in general form with Poisson brackets. $H_0$ denotes the Hamiltonian 
of the stellar orbits in the unperturbed disk. In order to obtain 
self--consistent perturbations the response density has to be inserted into 
the Poisson-equation
\begin{equation}
\Delta \delta\Phi = 4 \pi G \int d^2v f_1,
\end{equation}
where $G$ is the constant of gravitation. Unfortunately, the disk response
to a single Fourier component of the potential perturbation (1) is not a
Fourier component of the general disk response. I follow therefore Kalnajs
(1971) and take scalar products,
\begin{equation}
\frac{1}{4\pi^2} \int dx \int dy\,e^{-i(k'_{\rm x x} + k'_{\rm y} y)} . . .\,,
\end{equation}
of both sides of the Poisson-equation with
conjugate basis functions of the Fourier transform.
Details of the evaluation of the quadratures, which are carried out using
action and angle variables are given in (Fuchs 2001). The result is that
the Poisson-equation is converted to an integral equation of Volterra-type,
which is equivalent to the integral equation given by Julian \&
Toomre (1966), although it is not formulated in shearing coordinates,
\begin{equation}
\Phi_{\rm {\bf k'},\omega} = \int_{-\infty}^{k'_{\rm x}} dk_{\rm x} {\mathcal
K}(k_{\rm x},k'_{\rm x},k'_{\rm y},\omega) 
\Phi_{{\rm k}_{\rm x},{\rm k}'_{\rm y},\omega},
\end{equation}
with a kernel $\mathcal K$ that can be expressed analytically (Fuchs
2001)\footnote{Positive wave numbers $k'_{\rm y}$ will be assumed in the 
following.}. By Fredholm discretization
equation (5) can be transformed into a set of algebraic equations.
The kernel $\mathcal K$ vanishes on the diagonal, so that the triangular 
coefficient matrix of this set of equations has an unity diagonal, implying a 
non-vanishing determinant.  Thus the homogenous integral
equation (5) has no eigensolutions, indicating that there exist in
the shearing sheet no -- except ringlike ($k'_{\rm y} = 0$) -- proper spiral
modes in the sense of rigidly rotating spatial patterns with well defined growth
rates. External potential perturbations or initial ($t=0$) density or velocity
perturbations of the basic state of the disk are represented by an 
inhomogeneous term in the integral
equation (5), or the corresponding integral equation for the surface density
\begin{equation}
\Sigma_{\rm {\bf k'},\omega} = \int_{-\infty}^{k'_{\rm x}} dk_{\rm x} {\mathcal
K}(k_{\rm x},k'_{\rm x},k'_{\rm y},\omega)
\Sigma_{{\rm k}_{\rm x},{\rm k}'_{\rm y},\omega} + r_{\rm{\bf k'},\omega}.
\end{equation}
The resolvent kernel ${\mathcal R}(k_{\rm x},k'_{\rm x},k'_{\rm y},\omega)$
of the inhomogeneous integral equation (6) can be obtained
as a Neumann series.  Solutions of equation (6) are then found
in a unique way by a convolution
of ${\mathcal R}$ with the inhomogeneous terms $r_{\rm\bf k'}$. Transforming
these solutions back from $\omega$-
to time-domain one can show that the resulting spatial pattern is a
superposition of `swinging' density waves, shearing with the general flow, but
exhibiting transient growth as they swing by until they finally decay.
Fig.~2, taken from Fuchs (1991), illustrates this for single plane sinusoidal
waves, $r_{\rm\bf k'} \propto \delta(k'_{\rm x} - k_{\rm x}^{in})$, all of the
same initial radial wave number $k_x^{in}$.
\begin{figure}
\centering
%
%
\caption{Amplitudes of swing amplified density waves with initial radial wave
   numbers $k_{\rm x}^{in} = -1$ (leading waves). Wave numbers are given in
   units of $k_{\rm crit} = \kappa^2/2\pi G\Sigma_0$, where $\kappa$ denotes the
   epicyclic frequency. Time is in units of $(2Ak_{\rm y})^{-1}$. The Toomre
   stability parameter is $Q=1.4$,
   and an Oort constant of $A = \Omega_0/2$ is assumed.}
\label{fig:2}       
\end{figure}
The resulting spatial pattern evolves then as
\begin{eqnarray}
\Sigma(x,y,t)  =  \delta (t) e^{i(k_{\rm x}^{in}x+k'_{\rm y}y)}
+  \tilde{\mathcal R}(k_{\rm x}^{in},k_{\rm x}^{in}+2Ak'_{\rm y}t)
e^{i[(k_{\rm x}^{in}+2Ak'_{\rm y}t)x+k'_{\rm y}y]}\,,
\end{eqnarray}
where the swinging around of the wave crests of the density waves is described 
by the growth of the effective radial wave number
$k_{\rm x}^{in}+2Ak'_{\rm y}t$ with time. As is well known, amplification is 
high for density waves, which are initially
leading, but low for initially trailing waves.

A visual impression of density waves growing from initial random fluctuations
of the surface density of the sheet is given in Fig 3., where snapshots of a
numerical simulation of the dynamical evolution of the shearing sheet are shown
(Fuchs, Dettbarn \& Tsuchiya, in preparation).
\begin{figure}
\centering
%
%
\caption{Snapshots of the growth of density waves in a numerical simulation of 
the dynamical evolution of the shearing sheet. The sheet was seeded initially 
with random noise as shown in the left panel. The right panel shows the sheet
after an elapsed time of one epicyclic period. Length unit is the critical wave
length $\lambda_{\rm crit}= 2 \pi/k_{\rm crit}$.}
\label{fig:3}       
\end{figure}

\index{shearing sheet in a live dark halo }
\section{The shearing sheet immersed in a live dark halo}
\label{sec:3}

The evolution of the distribution function of the disk stars in phase space is
described by the linearized Boltzmann equation
\begin{eqnarray}
&&
\frac{\partial f_{\rm d1}}{\partial t} + u \frac{\partial f_{\rm d1}}
{\partial x} + v \frac{\partial f_{\rm d1}}{\partial y} 
- \frac{\partial \Phi_{\rm d0}+\Phi_{\rm h0}}{\partial x}\frac{\partial
f_{\rm d1}}{\partial u} 
- \frac{\partial \Phi_{\rm d0}+\Phi_{\rm h0}}{\partial y}
\frac{\partial f_{\rm d1}}{\partial v} \nonumber \\ &&
 -\frac{\partial \Phi_{\rm d1}+\Phi_{\rm h1}}{\partial x}\frac{\partial
 f_{\rm d0}}{\partial u} 
  - \frac{\partial \Phi_{\rm d1}+\Phi_{\rm h1}}{\partial y}
\frac{\partial f_{\rm d0}}{\partial v} =0 \,,
\end{eqnarray}
where ($u$, $v$) are the velocity components corresponding to the $x$ and $y$
coordinates, respectively. Equation (8) has been derived from the 
general 6--dimensional Boltzmann equation assuming delta--function like
dependencies of the distribution function on the vertical $z$ coordinate and 
the vertical $w$ velocity component, respectively, and integrating the
Boltzmann equation with respect to them. A perturbation Ansatz of the form
\begin{equation}
f_{\rm d} = f_{\rm{d0}} + f_{\rm{d1}} \,, \; 
\Phi_{\rm d} = \Phi_{\rm{d0}} + \Phi_{\rm{d1}}\,, \; 
\Phi_{\rm h} = \Phi_{\rm{h0}} + \Phi_{\rm{h1}} 
\end{equation}
is chosen for the distribution function of the disk stars and the
gravitational potentials of the disk and the halo, respectively, and the 
Boltzmann equation (8) has been linearized accordingly. 

Similarly the linearized Boltzmann equation for the halo particles can be 
written as
\begin{eqnarray}
&&
\frac{\partial f_{\rm h1}}{\partial t} + u \frac{\partial f_{\rm h1}}
{\partial x} + v \frac{\partial f_{\rm h1}}{\partial y}
+ w \frac{\partial f_{\rm h1}}{\partial z} 
- \frac{\partial \Phi_{\rm d0}+\Phi_{\rm h0}}{\partial x}\frac{\partial
f_{\rm h1}}{\partial u}- 
\frac{\partial \Phi_{\rm d0}+\Phi_{\rm h0}}{\partial y}\frac{\partial
f_{\rm h1}}{\partial v} \nonumber \\ && -\frac{\partial \Phi_{\rm d0}+
\Phi_{\rm h0}}{\partial z}\frac{\partial f_{\rm h1}}{\partial w} 
-\frac{\partial \Phi_{\rm h1}+\Phi_{\rm d1}}{\partial x}
\frac{\partial f_{\rm h0}}{\partial u} 
 - \frac{\partial \Phi_{\rm h1}+\Phi_{\rm d1}}{\partial y}
\frac{\partial f_{\rm h0}}{\partial v}  \nonumber \\ &&  
 - \frac{\partial \Phi_{\rm h1}+\Phi_{\rm d1}}{\partial z}
\frac{\partial f_{\rm h0}}{\partial w} = 0 \,.
\end{eqnarray}
The choice of the dark halo model was lead by the following considerations. One
of the deeper reasons for the success of the infinite shearing sheet model to 
describe spiral density waves realistically is the rapid convergence of the 
Poisson integral in self--gravitating disks (Julian \& Toomre 1966). Consider, 
for example, the potential of a sinusoidal density perturbation
\begin{equation} 
\Phi(x,y) = - G \int_{-\infty}^{\infty} dx' \int_{-\infty}^{\infty} dy'
\frac{\Sigma_{10} \sin{(kx')}}{\sqrt{ (x-x')^2 + (y-y')^2}} \,,
\end{equation} 
\begin{equation} 
\Phi(x,y=0) = - 4G \Sigma_{10} \sin{(kx)} \lim_{ x_{\rm L} \to 
\infty} \frac{{\rm Si}(k x_{\rm L})}{k} = 
 - \frac{2 \pi G \Sigma_{10}\sin{(kx)}}{k}\,.
\end{equation}
The sine integral in equation (12) converges so rapidly that it reaches at
$k x_{\rm L} = \frac{\pi}{2}$ already 87\% of its asymptotic value. Thus
the ``effective range'' of gravity is about only a quarter of a wave length. The
shearing sheet models effectively patches of galactic disks of such size. The
wave lengths of density waves are of the order of the critical wave length 
\begin{equation} 
\lambda_{\rm crit} = \frac{2 \pi}{k_{\rm crit}} = 
\frac{4 \pi^2 G \Sigma_{\rm 0}}{\kappa^2}\,,
\end{equation} 
where $\kappa$ denotes the epicyclic frequency of the stellar orbits and 
$\Sigma_{\rm 0}$ the surface density of the disk. In the solar neighbourhood
in the Milky Way, for instance, the critical wave length is $\lambda_{\rm
crit}=5$ kpc. Thus it is reasonable to neglect over such length scales, like in
the shearing sheet model, the curvature of the mean circular orbits of the 
stars around the galactic center or the gradient of the surface density. The 
curvature of the stellar orbits due to the epicyclic motions of the stars, 
on the other hand, cannot be neglected and is indeed not neglected in the
shearing sheet model. The radial size of an epicycle is approximately given by
$\sigma_{\rm u}/\kappa$, where $\sigma_{\rm u}$ denotes the radial velocity
dispersion of the stars, and the ratio of epicycle size and critical wave 
length is given by
\begin{equation} 
\frac{\sigma_{\rm u}}{\kappa \lambda_{\rm crit}} = 0.085 Q
\end{equation} 
in terms of the Toomre (1964) stability parameter $Q$  which is typically
of the order of 1 to 2. Concurrent to these approximations I have assumed a
dark halo which is homogeneous in its unperturbed state. 
Accordingly the curvature of the
unperturbed orbits of the halo particles is neglected on the scales considered
here and the particles are assumed to be on straight--line orbits. The equations
of motion of the halo particles are the characteristics of the Boltzmann 
equation. In a homogeneous halo $\ddot{\bf x}=\nabla (\Phi_{\rm d0}+
\Phi_{\rm h0}) = 0$, and in accordance with this assumption I neglect the force
terms $\nabla \Phi_{\rm d0}$ and $\nabla \Phi_{\rm h0}$ in the Boltzmann 
equation (10). This simplifies its solution considerably. The disadvantage of 
such a model is that there are no higher--order resonances of the 
orbits of the halo particles with the density waves as described by Weinberg 
(1985) or observed in the high--resolution simulations by Athanassoula (2002, 
2003). However their effect was shown to be much less important than the main 
resonances of the particles with the density waves, which are properly 
described in the present model.  
\begin{figure}
\centering
%
%
\caption{Sketch of the disk and halo model.}
\label{fig:4}       
\end{figure}
\subsection{Halo dynamics}

The Boltzmann equation (10) can be viewed as a linear partial differential 
equation for the perturbation of the distribution function of the halo 
particles, $f_{\rm  h1}$, with inhomogeneities depending on the perturbations 
of the gravitational potentials of the disk and the halo, $\nabla 
\Phi_{\rm d1}$ and $\nabla \Phi_{\rm h1}$, respectively. Thus the equation 
can be solved for the disk and halo inhomogeneities separately and the 
solutions combined afterwards by superposition. It was shown by Fuchs (2004a)
that the Boltzmann equation with the inhomogeneity $\nabla \Phi_{\rm h1}$
describes just the Jeans instability of the dark halo.
However, dark halos are thought to be dynamically hot systems and their Jeans 
lengths will be of the order of the size of the halos themselves. Thus this 
part of the solution of the Boltzmann equation (10) is uninteresting in the 
present context and will be not considered in the following.

More interesting is the remaining part of the Boltzmann equation (10),
\begin{eqnarray}
&&
\frac{\partial f_{\rm h1}}{\partial t} + u \frac{\partial f_{\rm h1}}
{\partial x} + v \frac{\partial f_{\rm h1}}{\partial y}
+ w \frac{\partial f_{\rm h1}}{\partial z} \\ &&
-\frac{\partial \Phi_{\rm d1}}{\partial x}\frac{\partial f_{\rm h0}}
{\partial u} \nonumber  - \frac{\partial \Phi_{\rm d1}}{\partial y}
\frac{\partial f_{\rm h0}}{\partial v} 
 - \frac{\partial \Phi_{\rm d1}}{\partial z}
\frac{\partial f_{\rm h0}}{\partial w} = 0 \nonumber \,,
\end{eqnarray}
which describes the halo response to a perturbation in the disk. If the
gravitational potential perturbation of the disk is Fourier expanded
the Fourier terms have the form (cf. equation 33 of Fuchs 2001)
\begin{equation}
\Phi_{\rm d{\bf k}_{||}}e^{i(\omega t + k_{\rm x} x + k_{\rm y} y) -
 k_{\rm ||}|z|}
\end{equation}
with $k_{||}=|{\bf k}_{||}| =\sqrt{k_{\rm x}^2 + k_{\rm y}^2} $. This can be
converted to Fourier coefficients of the halo potential in 3--dimensional 
${\bf k}$--space as
\begin{equation}
\Phi_{\rm d{\bf k}} = \frac{1}{2 \pi} \int_{-\infty}^{\infty} dz
\Phi_{\rm d{\bf k}_{||}} e^{-i k_{\rm z} z - k_{\rm ||}|z|} = \frac{1}{\pi}
\frac{k_{||}}{k_{||}^2 + k_{\rm z}^2}\Phi_{\rm d{\bf k}_{||}} \,.
\end{equation}
Notice that the coordinate $y$, which is  defined in the reference
system of the disk, is related to the $y$ coordinate in the reference system of
the halo due to the motion of the center of the shearing sheet as
\begin{equation}
y \rightarrow y -r_{\rm 0} \Omega_{\rm 0} t \,.
\end{equation}
The further solution of the Boltzmann equation is straightforward and is 
described in full detail in Fuchs (2004a). The distribution function is then 
integrated over velocity space to obtain the Fourier coefficients $\rho_{\rm
h{\bf k}}$ of the density perturbation of the dark halo. Next the gravitational
potential associated with this density distribution is calculated from the 
Poisson equation,
\begin{equation}
- k^2 \Phi_{\rm h{\bf k}} = 4 \pi G \rho_{\rm h{\bf k}}\,.
\end{equation}
Since the gravitational forces in equation (8) have to be
taken at the midplane $z = 0$, it is necessary to convert the solution of the
Boltzmann equation $\Phi_{\rm h{\bf k}}$ from the representation in {\bf k} 
space to a mixed representation in $({\bf k}_{||}, z)$ space leading to two
contributions
\begin{eqnarray}
&& \Phi_{\rm h{\bf k}_{||}}^{\rm nr}(z=0) = 
\int_{-\infty}^{\infty} d k_{\rm z}
\frac{k_{||}}{(k_{||}^2+ k_{\rm z}^2)^2} \Phi_{\rm d{\bf k}}\frac{4 G 
\rho_{\rm b}}{\sigma_{\rm h}^2}  \\ && \times \{ 1 + i \sqrt{\pi} \, 
\frac{k_{\rm y}r_{\rm 0} \Omega_{\rm 0}-\omega}{\sqrt{2} k \sigma_{\rm h}} \, 
{\rm erf} \left(i\frac{k_{\rm y}r_{\rm 0} \Omega_{\rm 0}-\omega}
{\sqrt{2} k \sigma_{\rm h}}\right) 
\exp{-\frac{(k_{\rm y}r_{\rm 0} 
\Omega_{\rm 0}- \omega)^2}{2 k^2 \sigma_{\rm h}^2}}\} \nonumber 
\end{eqnarray}
due to halo particles not in resonance with the potential perturbation and 
\begin{eqnarray}
&& \Phi_{\rm h{\bf k}_{||}}^{\rm res}(z=0) = - \int_{-\infty}^{\infty} 
d k_{\rm z} \frac{k_{||}}{(k_{||}^2+ k_{\rm z}^2)^2} \Phi_{\rm d{\bf k}}
i \sqrt{\pi} \frac{4 G \rho_{\rm b}}
{\sigma_{\rm h}^2} \nonumber \\ && \times  
\frac{\omega - k_{\rm y}r_{\rm 0} \Omega_{\rm 0}}{\sqrt{2} k \sigma_{\rm h}} \, 
\exp{-\frac{(k_{\rm y}r_{\rm 0} 
\Omega_{\rm 0}- \omega)^2}{2 k^2 \sigma_{\rm h}^2}} 
\end{eqnarray}
due to non--resonant halo particles. $\rho_{\rm b}$ and $\sigma_{\rm h}$ denote
the mean spatial density and the velocity dispersion of the halo particles,
respectively. The final result can be formally written as
\begin{equation}
\Phi_{\rm h{\bf k}_{||}}(z=0) =\Upsilon (\omega - k_{\rm y}r_{\rm 0} 
\Omega_{\rm 0}, k_{||}) \Phi_{\rm d{\bf k}_{||}} \,,
\end{equation}
where the real and imaginary parts of $\Upsilon$ are defined by equations (20)
and (21), respectively. Thus for any given frequency there is a contribution
both from the non--resonant and the resonant halo particles. 

\subsection{Disk dynamics} 

The halo response (22) to the perturbation in the disk has to be inserted into
equation (8). Solving the Boltzmann equation (8) is greatly facilitated by the
fact that its form is identical to the case of an isolated shearing sheet with
the replacement
\begin{equation}
\Phi_{\rm d{\bf k}} \rightarrow (1+\Upsilon)\Phi_{\rm d{\bf k}}
\end{equation}
and one can use directly the results of Fuchs (2001) even though the Boltzmann
equation is treated there using action and angle variables instead of the
Cartesian coordinates as in equation (8). In particular the factor
$1 + \Upsilon$ is carried straightforward through to the fundamental Volterra
integral equation (equation 68 of Fuchs 2001)
\begin{equation}
\Phi_{\rm {\bf k'},\omega} = \int_{-\infty}^{k'_{\rm{x}}} dk_{\rm{x}} 
\mathcal{K} 
\left(k_{\rm{x}},k'_{\rm{x}}\right) (1+\Upsilon(k_{\rm x},k'_{\rm y},\omega))
\Phi_{\rm k_{\rm x}, k'_{\rm y}, \omega} 
 +  r_{\rm{\bf{k'},\omega}} \,,
\end{equation}
where the kernel $\mathcal{K}$ is given by equation (67) of Fuchs (2001).  
$r_{\rm{\bf{k'}}}$ describes an inhomogeneity of equation (24) related  
to an initial non--equilibrium state of the shearing sheet. Equation (24) is
separating in the circumferential wave number $k'_{\rm y}$. 
In equation (24) the wave numbers are expressed in units of the critical wave
number $k_{\rm crit}$. This implies that the Volterra equation
describing a shearing sheet embedded in a rigid halo potential is formally the
same as that of an isolated shearing sheet, because in this case $\Upsilon = 0$
and the halo mass affects only the numerical values of the critical wave number 
$k_{\rm crit}$ and the stability parameter $Q$. It is
advantageous to consider equation (24) transformed back from frequency to time
domain. Splitting off the $\omega$--dependent term $\exp{i \omega  
\frac{k_{\rm x}-k'_{\rm x}}{2 A k'_{\rm y}}}$ from the kernel and making use 
of the convolution theorem of the Fourier transform of products of two
functions leads to
\begin{eqnarray}
&& \Phi_{\rm {\bf k'},t}  = \int_{-\infty}^{k'_{\rm{x}}} dk_{\rm{x}} 
\tilde{\mathcal{K}}\left(k_{\rm{x}},k'_{\rm{x}}\right)  \{
\int_{0}^{\infty} dt' \Phi_{\rm k_{\rm x}, k'_{\rm y},t'} \delta
\left( t-t'+ \frac{k_{\rm x}-k'_{\rm x}}{2 A k'_{\rm y}} \right) \nonumber\\ &&
+\int_{0}^{\infty} dt' \Phi_{\rm k_{\rm x}, k'_{\rm y}, t'} \mathcal{F}
\left(\Upsilon(k_{\rm x},k'_{\rm y},\omega)
e^{i\omega\frac{k_{\rm x}-k'_{\rm x}}{2 A k'_{\rm y}}}\right)_{\rm t-t'} \}
 +  r_{\rm {\bf k'},t}\,,
\end{eqnarray}
where the operator $\mathcal{F}$ denotes the Fourier transform from $\omega$ 
to time domain. In equation (33) I have assumed an initial perturbation of the 
disk at time $t = 0$ so that $\Phi_{{\rm k_{\rm x}, k'_{\rm y}},t'<0}=0$. The 
Fourier transform $\mathcal{F}$ is given by equation (34) of Fuchs (2004a).
If this is inserted into equation (25) it takes the form
\begin{eqnarray}
 && \Phi_{\rm {\bf k'},t} = \int_{-\infty}^{k'_{\rm{x}}} dk_{\rm{x}} 
\tilde{\mathcal{K}}\left(k_{\rm{x}},k'_{\rm{x}}\right) \{
\Phi_{\rm k_{\rm x}, k'_{\rm y},t+\frac{k_{\rm x}-k'_{\rm x}}{2 A k'_{\rm y}}}
 \\ && +
\int_{0}^{t+\frac{k_{\rm x}-k'_{\rm x}}{2 A k'_{\rm y}}} dt'
 \Phi_{\rm k_{\rm x}, k'_{\rm y}, t'} \mathcal{F}
\left(\Upsilon(k_{\rm x},k'_{\rm y},\omega)
e^{i\omega\frac{k_{\rm x}-k'_{\rm x}}{2 A k'_{\rm y}}}\right)_{\rm t-t'} \}
+ r_{\rm {\bf k'},t} \,. \nonumber 
\end{eqnarray}
Equation (26) can be integrated numerically with very modest numerical effort.
In Fig.~5 I illustrate the response of the shearing sheet now embedded in a live
halo to an initial sinusoidal perturbation of unit amplitude. For this purpose 
I use the inhomogeneity term of the Volterra equation
\begin{equation}
r_{\rm {\bf k'},\omega} = \int_{-\infty}^{k'_{\rm{x}}} dk_{\rm{x}} 
{\mathcal{L}}\left(k_{\rm{x}},k'_{\rm{x}}\right)
f_{\rm k_{\rm x}, {k'}_{\rm y}}(0)
\end{equation}
derived in Fuchs (2001) with $f_{\rm k_{\rm x}, {k'}_{\rm y}}(0) 
\propto \delta ( k_{\rm x} -k_{\rm x}^{\rm in}) $. The response of the 
shearing sheet to this initial impulse is a swing amplification event as
described in section (2). The radial wave number $k_{\rm x}$ evolves as
\begin{equation}
k_{\rm x}=k_{\rm x}^{\rm in} +2 A {k'_{\rm y}} t \,,
\end{equation}
while the circumferential wave number ${k'_{\rm y}}$ is constant,
which means that the wave crests swing around from leading to trailing 
orientation during the amplification phase. Around $t = 6$ the amplitudes 
become negative which indicates that the swing amplified density wave is also 
oscillating. As can be seen from Fig.~5 comparing the evolution of 
shearing sheets embedded either in a rigid halo potential or a live halo this
characteristic behaviour of the density wave is not changed by the responsive 
halo, but the maximum growth factor of the amplitude of the wave is enhanced by
a surprisingly large amount. 
\begin{figure}
\centering
%
%
\caption{Swing amplified density wave in the shearing sheet. The upper diagram
shows the evolution in a shearing sheet embedded 
in a static halo potential triggered by an impulse with 
unit amplitude and wave vector ${\bf k}^{\rm in}=(-2,0.5)k_{\rm crit}$.  
Time is given in units of $(2Ak^{\rm in}_{\rm y}/k_{\rm crit})^{-1}$. 
The middle diagram shows the
evolution of a shearing sheet embedded in a live dark halo triggered by the
same impulse. The lower diagram shows the difference. The model parameters are
$A/\Omega_0=0.5, Q=1.4, \sigma_{\rm d}: \sigma_{\rm h}=1:5, G\rho_{\rm
b}/\kappa^2=0.01$, and $r_0\Omega_0:\sigma_{\rm d}=220:44$.}
\label{fig:5}       
\end{figure}

The enhanced maximum growth factor of swing amplified density
waves due to a responsive halo seems to be
the equivalent of the enhanced growth of bars of stellar disks embedded in live
dark halos seen in the numerical simulations. However, the interaction of the 
shearing sheet and the surrounding halo is not only mediated by the resonant 
halo particles, but the non--resonant halo particles play an important role as
well.The amplification of density waves depends critically on the Toomre 
stability
parameter $Q$. This is illustrated in Fig.~6 where the response of the shearing
sheet to the same initial impulse as in the previous example is shown, but
assuming a stability parameter of $Q=2$. As can be seen in Fig.~6 there is
neither effective amplification of density waves in a shearing sheet in a rigid
halo potential or in a shearing sheet embedded in a live dark halo.
\begin{figure}
\centering
%
%
\caption{Same as in Fig.~5, but adopting $Q=2$.}
\label{fig:6}       
\end{figure}

\index{wakes in dark halos}
\index{wakes in dark matter halos}
\section{Wakes in dark matter halos}
\label{sec:4}

The perturbations of the gravitational potential and the surface density of the
shearing sheet have their counterparts in the dark matter halo. From equation
(19) one obtains
\begin{equation}
\rho_{\rm h {\bf k}||}(z,\omega)= - \int_{-\infty}^\infty d k_{\rm z}
\frac{k^2}{4 \pi G} \Phi_{\rm h {\bf k}} e^{ik_{\rm z}z} \,,
\end{equation} 
where the Fourier coefficients $\Phi_{\rm h {\bf k}}$ derived in section (3.1)
have to be inserted,
\begin{eqnarray}
&& \rho_{\rm h{\bf k}_{||}}(z,\omega) = 
-  \int_{-\infty}^{\infty} d k_{\rm z}
\frac{k_{||}}{k_{||}^2+k_{\rm z}^2} \Phi_{\rm d{\bf k||}}\frac{ 
\rho_{\rm b}}{ \pi \sigma_{\rm h}^2} e^{i k_{\rm z}z}
 \\ && \times \{ 1 + i \sqrt{\pi} \, 
\frac{k_{\rm y}r_{\rm 0} \Omega_{\rm 0}-\omega}{\sqrt{2} k \sigma_{\rm h}} \, 
\left( 1 +{\rm erf} \left(i\frac{k_{\rm y}r_{\rm 0} \Omega_{\rm 0}-\omega}
{\sqrt{2} k \sigma_{\rm h}}\right) \right)
\exp{-\frac{(k_{\rm y}r_{\rm 0} 
\Omega_{\rm 0}- \omega)^2}{2 k^2 \sigma_{\rm h}^2}}\} \,.
\nonumber 
\end{eqnarray}
Equation (30) can be Fourier transformed back from frequency to time domain
in the same way as equation (32) of Fuchs (2004a) was Fourier transformed back 
to time domain  by a convolution of $\Phi_{\rm d{\bf k}||}$(t') with the 
Fourier transform of the remaining terms under the integral with respect to 
$k_{\rm z}$. To these terms the operator
\begin{equation}
\int_{-\infty}^\infty d \omega e^{i(\omega-k_{\rm y} r_0 \Omega_0) t} ...
\end{equation} 
is applied and the integral over $\omega$ is evaluated with the help of
formulae (6.317) and (3.952) of Gradshteyn \& Ryzhik (2000) leading effectively
to expression (34) of Fuchs (2004a) with $t-t' +\frac{k_{\rm x}-k'_{\rm x}}
{2 A k'_{\rm y}}$ replaced by $t-t'$ only. The integral over $k_{\rm z}$ can be
then calculated using formulae (3.723) and (3.896) of Gradshteyn \& Ryzhik 
(2000) with the result
\begin{eqnarray}
&& \rho_{\rm h {\bf k}||}(z,t) = - \frac{\rho_{\rm b}}{\sigma_{\rm h}^2}
\{ 2 \pi \Phi_{\rm d{\bf k}||}(t) e^{-k_{||}|z|} \\ &&
 +\sqrt{8 \pi}
\sigma_{\rm h} k_{||} \int_{-\infty}^t dt' \Phi_{\rm d{\bf k}||}(t')
e^{-\frac{1}{2}\sigma_{\rm h}^2 k_{||}^2 (t-t')^2} e^{i k_{\rm y} r_0 \Omega_0)
(t-t')} e^{- \frac{z^2}{2 \sigma_{\rm h}^2(t-t')^2}} \} \nonumber \,,
\end{eqnarray}
and finally I convert the potential perturbation of the disk to
the perturbation of its surface density with the relation
$k_{||}\Phi_{\rm d{\bf k}||}=-2\pi G \Sigma_{\rm d{\bf k}||}$ (cf.~Fuchs 2001)
and obtain
\begin{eqnarray}
&&\frac{\rho_{\rm h {\bf k}||}}{\rho_{\rm b}} = \frac{2 \pi G \Sigma_{\rm d}}
{\sigma_{\rm h}^2}\{ \frac{2 \pi}{k_{||}} \frac{\Sigma_{\rm d{\bf k}||}(t)}
{\Sigma_{\rm d}} e^{-k_{||}|z|} \\ && +\sqrt{8 \pi}\sigma_{\rm h}
\int_{-\infty}^t dt'\frac{\Sigma_{\rm d{\bf k}||}(t')}{\Sigma_{\rm d}}
e^{-\frac{1}{2}\sigma_{\rm h}^2 k_{||}^2 (t-t')^2} e^{i k_{\rm y} r_0 \Omega_0)
(t-t')} e^{- \frac{z^2}{2 \sigma_{\rm h}^2(t-t')^2}} \}\,. \nonumber 
\end{eqnarray}
The time integral in equation (33) has been calculated numerically. At the time
of maximal amplification of the density wave I find, adopting the same
parameters used to calculate Fig.~5, 
$\rho_{\rm h{\bf k}}/\rho_{\rm b}|_{\rm peak}= 0.6\,
\Sigma_{\rm d{\bf k}}/\Sigma_0|_{peak}$. Fig.~7 shows the vertical profile
of the density perturbation of the dark halo according to equation (33). As can
be seen from Fig.~7 the profile indicates a density enhancement of dark halo
matter close above the density wave crest. At larger distances above the
midplane there is a density deficit above the density wave crest, which means
that the cloud of dark matter particles is lagging behind like a trail of smoke.
\begin{figure}
\centering
\includegraphics[height=6cm]{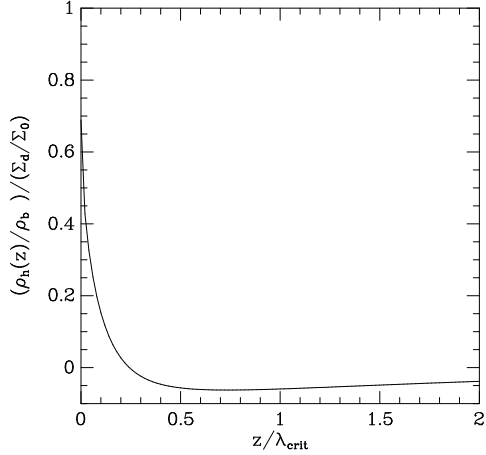}
\caption{Vertical profile of the density perturbation in the dark halo induced
by a density wave in the sheet, which is at maximal amplification, exactly 
above the density wave crest. The relative density contrast 
$\rho_{\rm h{\bf k}}/\rho_{\rm b}|_{\rm peak}$ is given in terms of the 
relative density contrast of the surface density of the sheet
$\Sigma_{\rm d{\bf k}}/\Sigma_0|_{\rm peak}$.}
\label{fig:7}       
\end{figure}

The existence of wakes in dark matter halos might have quite practical
implications. For instance the bar in the Milky Way focusses dark matter
particles dynamically into some regions in phase space and depopulates others, 
which may very well affect the flux of dark matter particles through detectors
on Earth. Such features have been observed as `star streams' in velocity space 
among the stars in the solar neighbourhood (Dehnen 2000, M\"uhlbauer \& Dehnen 
2003). These stars are, of course, stars members of the Milky Way disk, but 
similar effects are to be expected among halo objects. 
%

\begin{thebibliography}{99.}

   \bibitem{atha} E. Athanassoula:  Astroph. J., \textbf{569}, L83 (2002) 
   
   \bibitem{athana} E. Athanassoula:  Mon. Not. R. Astron. Soc., \textbf{341},
    1179 (2003)
  
   \bibitem{deba} V.P. Debattista, J. A. Sellwood: Astroph. J., \textbf{543},
    704 (2000)
    
   \bibitem{deh} W. Dehnen: Astron. J., \textbf{119}, 800 (2000)

   \bibitem{fuchs} B. Fuchs: Recurrent swing amplification induced by
   nonlinear amplitude effects. In: \textit{Dynamics of disc galaxies} 
   ed by B. Sundelius (G\"oteborgs Univ. and Chalmers Univ. of Tech., 
   G\"oteborg 1991) pp 359--363

   \bibitem{fuchsb} B. Fuchs: Astron. Astroph., \textbf{368}, 107 (2001) 
   
   \bibitem{fuc} B. Fuchs: Astron. Astroph., \textbf{419}, 941 (2004a)
   
   \bibitem{fu} B. Fuchs: Astron. Astroph., submitted (2004b)

   \bibitem{gold} P. Goldreich, D. Lynden--Bell: Mon. Not. R. Astron. Soc.,
    \textbf{130}, 125 (1965)
    
   \bibitem{grad} Gradshteyn, I.S., Ryzhik, I.M.: \textit{Table of 
   Integrals, Series, and Products}, 6th edn (Academic Press, New York 2000)

   \bibitem{julia} W.H. Julian, A. Toomre: Astroph. J., \textbf{146}, 810 
   (1966)
   
   \bibitem{kaln} A. Kalnajs: 1971, Astroph. J., \textbf{166}, 275 (1971)
   
   \bibitem{mude} G. M\"uhlbauer, W. Dehnen: Astron. Astroph., \textbf{401},
    975 (2003)

   \bibitem{ospe} J.P. Ostriker, P.J.E. Peebles: Astroph. J., \textbf{186}, 
   467 (1973)

  \bibitem{toom} A. Toomre: Astroph. J., \textbf{139}, 1217 (1964)
  
  \bibitem{toomr} A. Toomre: Ann. Rev. Astron. Astroph., \textbf{15},
   437 (1977)

   \bibitem{too} A. Toomre: What amplifies the spirals? 
   In: \textit{The Structure and Evolution of Normal Galaxies} 
   ed by S.M. Fall, D. Lynden--Bell (Cambridge Univ. Press,
   Cambridge 1981) pp 111--136
   
   \bibitem{wei} M.D. Weinberg: Mon. Not. R. Astron. Soc., \textbf{213},
    451 (1985)
   
\end{thebibliography}
%


\end{document}